\documentclass{emulateapj}
\usepackage{apjfonts}
\usepackage[]{natbib}
\usepackage{graphics}

\newcommand{\etal}{et al.}  
\newcommand{\per}{\ensuremath{^{-1}}}
\newcommand{\persq}{\ensuremath{^{-2}}}

\newcommand{\hal}{H\ensuremath{\alpha}}
\newcommand{\hbeta}{H\ensuremath{\beta}} 
\newcommand{\hgamma}{H\ensuremath{\gamma}} 
\newcommand{\hdelta}{H\ensuremath{\delta}} 

\newcommand{\msun}{\ensuremath{M_{\odot}}}
 
\newcommand{\kms}{km~s\ensuremath{^{-1}}}

\newcommand{\mbh}{\ensuremath{M_\mathrm{BH}}}

\newcommand{\msigma}{\ensuremath{\mbh-\sigma}}

\newcommand{\lbol}{\ensuremath{L_{\mathrm{bol}}}}

\newcommand{\kepler}{\emph{Kepler}}
\newcommand{\tpeak}{\ensuremath{\tau_\mathrm{peak}}}
\newcommand{\tcen}{\ensuremath{\tau_\mathrm{cen}}}
\newcommand{\rmax}{\ensuremath{R_\mathrm{max}}}
\newcommand{\fvar}{\ensuremath{F_\mathrm{var}}}
\newcommand{\sigmaline}{\ensuremath{\sigma_\mathrm{line}}}
\newcommand{\sigmainst}{\ensuremath{\sigma_\mathrm{inst}}}
\newcommand{\eddratio}{\ensuremath{L/L_\mathrm{Edd}}}

\shorttitle{REVERBERATION IN ZW 229-015}
\shortauthors{BARTH ET AL.}

\begin{document} 

\title{Broad-Line Reverberation in the Kepler-Field Seyfert Galaxy
  Zw 229-015}

\author{
  Aaron J. Barth\altaffilmark{1}, 
  My L. Nguyen\altaffilmark{1},
  Matthew A. Malkan\altaffilmark{2}, 
  Alexei V. Filippenko\altaffilmark{3},
  Weidong Li\altaffilmark{3}, 
  Varoujan Gorjian\altaffilmark{4},
  Michael D. Joner\altaffilmark{5}, 
  Vardha Nicola Bennert\altaffilmark{6},
  Janos Botyanszki\altaffilmark{7},
  S. Bradley Cenko\altaffilmark{3},
  Michael Childress\altaffilmark{7},
  Jieun Choi\altaffilmark{3},
  Julia M. Comerford\altaffilmark{8},
  Antonino Cucciara\altaffilmark{7},
  Robert da Silva\altaffilmark{9},
  Gaspard Duch\^{e}ne\altaffilmark{3,10},
  Michele Fumagalli\altaffilmark{11},
  Mohan Ganeshalingam\altaffilmark{3},
  Elinor L. Gates\altaffilmark{12},
  Brian F. Gerke\altaffilmark{13}, 
  Christopher V. Griffith\altaffilmark{14},
  Chelsea Harris\altaffilmark{6}, 
  Eric G. Hintz\altaffilmark{5}, 
  Eric Hsiao\altaffilmark{7},
  Michael T. Kandrashoff\altaffilmark{3},
  William C. Keel\altaffilmark{15}, 
  David Kirkman\altaffilmark{16},
  Io K. W. Kleiser\altaffilmark{3},
  C. David Laney\altaffilmark{5}, 
  Jeffrey Lee\altaffilmark{16},
  Liliana Lopez\altaffilmark{16},
  Thomas B. Lowe\altaffilmark{12},
  J. Ward Moody\altaffilmark{5}, 
  Alekzandir Morton\altaffilmark{3},
  A. M. Nierenberg\altaffilmark{6}, 
  Peter Nugent\altaffilmark{3,7},
  Anna Pancoast\altaffilmark{6}, 
  Jacob Rex\altaffilmark{3},
  R. Michael Rich\altaffilmark{2}, 
  Jeffrey M. Silverman\altaffilmark{3},
  Graeme H. Smith\altaffilmark{9},
  Alessandro Sonnenfeld\altaffilmark{6}, 
  Nao Suzuki\altaffilmark{7},
  David Tytler\altaffilmark{16},
  Jonelle L. Walsh\altaffilmark{1}, 
  Jong-Hak Woo\altaffilmark{17},
  Yizhe Yang\altaffilmark{18}, and
  Carl Zeisse\altaffilmark{16}
}

\altaffiltext{1}{Department of Physics and Astronomy, 4129 Frederick
  Reines Hall, University of California, Irvine, CA, 92697-4575, USA;
  barth@uci.edu}

\altaffiltext{2}{Department of Physics and Astronomy, University of
California, Los Angeles, CA 90024, USA}

\altaffiltext{3}{Department of Astronomy, University of California,
Berkeley, CA 94720-3411, USA}

\altaffiltext{4}{Jet Propulsion Laboratory, 4800 Oak Grove Boulevard,
MS 169-327, Pasadena, CA 91109, USA}

\altaffiltext{5}{Department of Physics and Astronomy, N283 ESC,
Brigham Young University, Provo, UT 84602-4360, USA}

\altaffiltext{6}{Department of Physics, University of California,
Santa Barbara, CA 93106, USA}

\altaffiltext{7}{Lawrence Berkeley National Laboratory, 1 Cyclotron
Road, Berkeley, CA 94720, USA}

\altaffiltext{8}{Astronomy Department, University of Texas at Austin,
Austin, TX 78712, USA}

\altaffiltext{9}{University of California Observatories/Lick
  Observatory, Santa Cruz, CA 95064, USA}

\altaffiltext{10}{UJF-Grenoble 1 / CNRS-INSU, Institut de
Plan\'etologie et d'Astrophysique de Grenoble (IPAG) UMR 5274,
Grenoble, F-38041, France}

\altaffiltext{11}{Department of Astronomy and Astrophysics, University
of California, Santa Cruz, CA 95064, USA}

\altaffiltext{12}{Lick Observatory, P.O. Box 85, Mount Hamilton, CA
  95140, USA}

\altaffiltext{13}{Kavli Institute for Particle Astrophysics and
Cosmology, SLAC National Accelerator Laboratory, 2575 Sand Hill Rd.,
M/S 29, Menlo Park, CA 94025, USA}

\altaffiltext{14}{Department of Astronomy and Astrophysics, The
Pennsylvania State University, 525 Davey Lab, University Park, PA
16802, USA}

\altaffiltext{15}{Department of Physics and Astronomy, University of
Alabama, P.O. Box 870324, Tuscaloosa, AL 35487, USA}

\altaffiltext{16}{Center for Astrophysics and Space Sciences,
  University of California, San Diego, CA 92093-0424, USA}

\altaffiltext{17}{Astronomy Program, Department of Physics and
Astronomy, Seoul National University, Seoul, Korea, 151-742}

\altaffiltext{18}{Department of Physics and Department of Mathematics,
University of California, Berkeley, CA 94720, USA}

\slugcomment{To appear in The Astrophysical Journal}

\begin{abstract}

The Seyfert 1 galaxy Zw 229-015 is among the brightest active galaxies
being monitored by the \kepler\ mission.  In order to determine the
black hole mass in Zw 229-015 from \hbeta\ reverberation mapping, we
have carried out nightly observations with the Kast Spectrograph at
the Lick 3~m telescope during the dark runs from June through December
2010, obtaining 54 spectroscopic observations in total.  We have also
obtained nightly $V$-band imaging with the Katzman Automatic Imaging
Telescope at Lick Observatory and with the 0.9~m telescope at the
Brigham Young University West Mountain Observatory over the same
period.  We detect strong variability in the source, which exhibited
more than a factor of 2 change in broad \hbeta\ flux.  From
cross-correlation measurements, we find that the \hbeta\ light curve
has a rest-frame lag of $3.86^{+0.69}_{-0.90}$ days with respect to
the $V$-band continuum variations.  We also measure reverberation lags
for \hal\ and \hgamma\ and find an upper limit to the \hdelta\ lag.
Combining the \hbeta\ lag measurement with a broad \hbeta\ width of
$\sigmaline = 1590 \pm 47$ \kms\ measured from the root-mean-square
variability spectrum, we obtain a virial estimate of $\mbh =
1.00_{-0.24}^{+0.19} \times10^7$ \msun\ for the black hole in Zw
229-015.  As a \kepler\ target, Zw 229-015 will eventually have one of
the highest-quality optical light curves ever measured for any active
galaxy, and the black hole mass determined from reverberation mapping
will serve as a benchmark for testing relationships between black hole
mass and continuum variability characteristics in active galactic
nuclei.

\end{abstract}


\section{Introduction}

The NASA \kepler\ mission \citep{borucki2010a} is designed to search
for exoplanets by detecting transit events, and early results have
demonstrated the spectacular quality of \kepler's ultra-precise
photometry \citep[e.g.,][]{borucki2010b}.  In addition to its primary
goal of obtaining light curves of Galactic stars, \kepler\ is
monitoring a small number of active galactic nuclei (AGNs) of various
types, including Seyfert galaxies, quasars, and blazars.  \kepler\
will provide the best optical light curves ever obtained of AGNs,
vastly exceeding the quality of any previous observations in terms of
both sampling cadence and photometric accuracy.  This will enable
unprecedented measurements of the characteristic properties of AGN
optical variability.

One of the key motivations for measurement of high-precision AGN light
curves is to determine the power spectrum of continuum fluctuations
and to search for any characteristic timescales of variability (in the
form of breaks or slope changes in the power spectrum of variations),
which might be related to light-crossing, dynamical, or thermal
timescales in the accretion disk.  At X-ray energies, long-duration
light curves have been used to search for breaks in the power spectral
density \citep[e.g.,][]{uttley2002, markowitz2003}, and the
break timescales are observed to be dependent on both black hole mass
and luminosity across a range extending from X-ray binaries to quasars
\citep{mchardy2006}.  There is accumulating evidence that the optical
fluctuations of AGNs also contain characteristic timescales that may
depend on black hole mass and luminosity
\citep{collier2001, kelly2009,mcleod2010}.  

Ground-based optical light curves of AGNs are limited in quality by
nightly and seasonal sampling, weather, and seeing-dependent errors,
but \kepler\ can provide exquisite data that will for the first time
enable detailed measurements of the optical variability properties of
individual objects over timescales ranging from hours up to
(potentially) a few years.  Unobscured (Type 1) AGNs monitored by
\kepler\ can provide powerful tests of proposed relationships between
variability properties, black hole mass, and luminosity that have been
derived from larger samples of objects with lower-quality ground-based
light curves.

In order to fully exploit the results of these \kepler\ AGN
observations, independent measurements of black hole masses in
\kepler-field AGNs are needed.  Here, we present the results of a
six-month, ground-based, optical reverberation-mapping program for the
\kepler\ target Zw 229-015 \citep[at $z=0.0275$;][]{falco1999}, with
the goal of determining the black hole mass in this galaxy from its
\hbeta\ reverberation lag.

This galaxy (also known as CGCG 229-015 or 2MASX J19052592+4227398)
was first identified as a Type 1 Seyfert by \citet{proust1990}.  Since
then, no follow-up observations of this AGN have been reported in the
literature at all, aside from its inclusion in \emph{ROSAT} All-Sky
Survey source catalogs \citep{voges1999,zimmermann2001}.
Nevertheless, Zw 229-015 is one of the nearest and brightest AGNs in
the \kepler\ field, and it was selected as a monitoring target for the
first two cycles of \kepler\ observations.  To assess its suitability
for ground-based monitoring, we obtained a test spectroscopic exposure
at the Lick Observatory 3~m Shane telescope in April 2010.  This
observation reconfirmed the Seyfert 1 classification, and we then
selected it for the reverberation mapping program reported here.  In
this paper, we describe our imaging and spectroscopic observations of
Zw 229-015, the measurements of the continuum and emission-line light
curves and the broad emission-line reverberation lag, and an estimate
of the black hole mass based on the \hbeta\ lag and line width.

\section{Imaging Observations}

Reverberation measurements require a light curve of the AGN
continuum, and we obtained $V$-band photometry of Zw 229-015 following
the same strategy used in recent reverberation programs
\citep[e.g.,][]{walsh2009, denney2010}.  Although ground-based
measurements cannot approach the accuracy or sampling cadence of the
\kepler\ data, they allowed us to monitor the target's variability
continuously during the course of our observing program.  Furthermore,
the $V$-band photometry gives a nearly line-free measurement of the
galaxy's continuum flux, while the broad \kepler\ spectral response
includes the \hal\ and \hbeta\ lines from the AGN.

\begin{figure}
\plotone{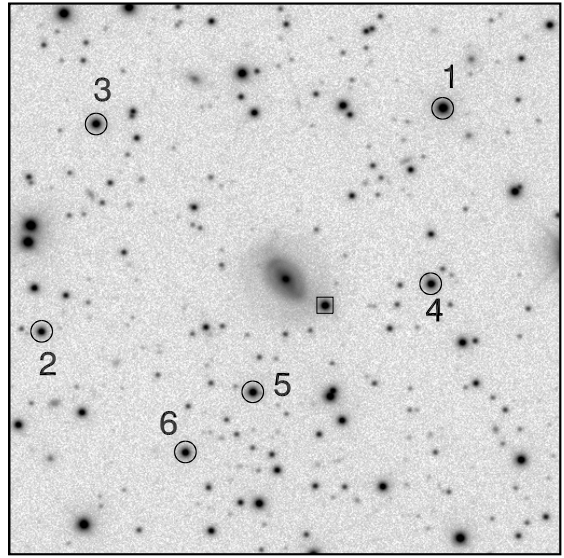}
\caption{$V$-band image centered on Zw 229-015, created by stacking
  the 18 best-seeing frames (all with seeing better than 1\farcs5)
  from the West Mountain Observatory.  North is up and east to the
  left, and the size of the displayed region is
  $250\arcsec \times 250\arcsec$.  Photometric comparison stars are
  circled and labeled, and the star that fell in the spectroscopic
  aperture is marked with a square.
\label{wmoimage}}
\end{figure}

We obtained regularly scheduled $V$-band imaging at the 0.76~m Katzman
Automatic Imaging Telescope (KAIT) at Lick Observatory, and at the
0.9~m telescope at the Brigham Young University West Mountain
Observatory (WMO).  At both telescopes, Zw 229-015 was observed each
night during the monitoring period, except when precluded by weather
or instrument problems.  We also obtained a small number of images at
other facilities, as described below.  Figure \ref{wmoimage}
illustrates the field surrounding Zw 229-015.  The galaxy has a barred
spiral morphology, with a Hubble type of roughly SBa.

\subsection{KAIT}

The KAIT 0.76~m robotic telescope \citep{filippenko2001} uses a
$512 \times 512$ pixel SITe CCD camera with a scale of 0\farcs8
pixel\per.  A single 300~s exposure of Zw 229-015 was obtained on each
night, weather permitting, from June 3 through December 13 (UT dates
are used throughout this paper). During the
monitoring period, KAIT had two month-long shutdowns for repairs,
first in August and again from mid-October to mid-November.  The
median seeing for all KAIT exposures was 2\farcs15.  The automatic
KAIT data-processing pipeline includes bias subtraction and
flatfielding.

\subsection{West Mountain Observatory}

The 0.9~m telescope at the West Mountain Observatory is one of five
small research telescopes operated by Brigham Young University.  It
was installed and tested in late 2009 and began working in a partially
queue-scheduled mode during the 2010 observing season.  The
observations secured for this investigation were scheduled as part of
an external guest investigator program.

From June 5 until September 1, observations were made using a
$3056 \times 3056$ pixel KAF 09000 CCD camera with a scale of 0\farcs5
pixel\per.  After September 2, observations were made using a
$2048 \times 2048$ pixel Fairchild 3041-UV CCD camera with a scale of
0\farcs6 pixel\per.  The typical observing procedure was to secure two
exposures of Zw 229-015 while it was close to meridian transit on each
night that conditions permitted telescope operations.  Exposures were
300~s from the start of monitoring until September 7, and were
decreased to 200~s thereafter because of the higher sensitivity of the
Fairchild CCD.  Observations continued until December 13, but only a
few exposures were obtained after November 5 due to poor weather.
Standard-star fields from \citet{landolt1992} were observed on a few
photometric nights. The median seeing for the WMO exposures was 1\farcs89.
The standard processing pipeline for WMO data includes bias
subtraction and flat-fielding.

\subsection{The Lick Nickel Telescope}

The Lick 1~m Nickel reflector was used to observe Zw 229-015 during a
few nights while KAIT was shut down for repairs.  The Nickel Direct
Imaging Camera contains a thinned, Loral $2048 \times 2048$ pixel CCD
with a scale of 0\farcs184 pixel\per, and the observations were
obtained using $2 \times 2$ binning for the CCD readout.  On one
photometric night, \citet{landolt1992} standard-star fields were also
observed.  Overscan subtraction and flat-fielding using twilight sky
flats were done following standard procedures in IRAF\footnote{IRAF is
distributed by the National Optical Astronomy Observatories, which are
operated by the Association of Universities for Research in Astronomy,
Inc., under cooperative agreement with the National Science
Foundation.}.

\subsection{Maidanak Observatory}

Zw 229-015 was observed on five nights in August and September using
the Seoul National University 4096 $\times$ 4096 pixel camera
\citep[SNUCAM;][]{im2010} and $V$-band filter at the Maidanak
Observatory 1.5~m telescope in Uzbekistan (Fairchild CCD; 0\farcs266
pixel\per). The processing pipeline for SNUCAM data included overscan
subtraction and flat-fielding.

\begin{deluxetable}{lccc}
\tablecaption{Photometric Comparison Stars}
\tablehead{
  \colhead{Star} &
  \colhead{$\alpha$} &
  \colhead{$\delta$} &
  \colhead{$V$} \\
  \colhead{} &
  \colhead{($h~m~s$)} &
  \colhead{($\circ~\prime~\prime\prime$)} &
  \colhead{(mag)}
}
\startdata
1 & 19:05:19.49 & 42:28:57.7 & $14.983\pm0.006$ \\
2 & 19:05:35.95 & 42:27:16.2 & $16.575\pm0.014$ \\
3 & 19:05:33.71 & 42:28:50.5 & $15.954\pm0.011$ \\
4 & 19:05:19.98 & 42:27:37.9 & $16.090\pm0.016$ \\
5 & 19:05:27.28 & 42:26:48.8 & $16.000\pm0.015$ \\
6 & 19:05:30.05 & 42:26:21.4 & $16.198\pm0.016$
\enddata
\tablecomments{Coordinates are J2000, and are based on an astrometric
  solution obtained using the astrometry.net package
  \citep{lang2010}. The quoted magnitude uncertainties are calculated
  as the standard deviation of all measurements from KAIT and WMO for
  the complete light curve of each comparison star.
 }
\label{compstarstable}
\end{deluxetable}

\section{Photometric Reductions and Measurements}
\label{photsection}

Our light-curve measurements followed the same procedures used by
\citet{walsh2009}.  First, each image was cleaned of cosmic-ray hits
using the LA-COSMIC procedure \citep{vandokkum2001} implemented in
IRAF.  Aperture photometry on Zw 229-015 and several comparison stars
was performed using the IRAF task \texttt{phot}.  Ultimately, six
comparison stars were selected; these are illustrated in Figure
\ref{wmoimage}.  The photometric measurements were carried out with a
range of aperture sizes. We found that the 4\arcsec\ aperture radius
gave the cleanest light curves, and our final photometry is based on
this aperture size.  The background sky annulus had inner and outer
radii of 8\arcsec\ and 14\farcs4.  When more than one exposure was
taken per night at a given telescope, the measurements were averaged
together to give a single data point for that night.

Photometric calibration was accomplished by carrying out aperture
photometry on the \citet{landolt1992} standard stars observed at WMO
and the Nickel telescope on nights that were judged to be photometric
by the observers.  We used only $V$-band exposures and did not attempt
to derive any color-dependent terms for the transformation between
instrumental and calibrated $V$ magnitudes.  We found excellent
agreement between the photometric calibrations for the night of
September 26 at WMO and the night of August 12 at the Nickel
telescope, and took the average of the two calibrations to obtain our
final photometric scale.  Table \ref{compstarstable} lists the
positions and magnitudes of the six comparison stars.

The light curves from each telescope were separately normalized to
$V$-band (Vega-system) magnitudes based on the comparison-star
photometry.  To check for small systematic differences between the
photometric scales of the four telescopes, which could be present
because color terms were not measured, we determined the average
offset between the KAIT and WMO light curves for the 53 nights when
the object was observed at both telescopes.  The weighted average
offset was $0.010 \pm 0.001$ mag, and we applied this shift of 0.01
mag to the KAIT light curve as an empirical correction to bring it
into agreement with the WMO photometric scale.  Similar offsets of
0.010 and 0.020 mag were applied to the Nickel and Maidanak data
points, respectively, in order to bring them into agreement with the
combined KAIT and WMO light curve, based on the weighted average shift
determined from nights when the galaxy was observed at more than one
telescope.

\begin{deluxetable}{lcc}
\tablecaption{Photometric Observations and Measurements}
\tablehead{
  \colhead{HJD} &
  \colhead{Telescope} &
  \colhead{$V$} \\
  \colhead{$-2450000$} &
  \colhead{} &
  \colhead{(mag)}
}
\startdata
    5351.9785  &  K  &    $15.989  \pm   0.008 $  \\
    5352.8036  &  W  &    $15.980  \pm   0.003 $  \\
    5352.9688  &  K  &    $15.976  \pm   0.007 $  \\
    5353.9834  &  K  &    $15.976  \pm   0.009 $  \\
    5354.9619  &  K  &    $15.957  \pm   0.008 $  \\
\enddata
\tablecomments{Telescopes are listed as follows: K = KAIT, W = WMO, N =
  Lick Nickel, M = Maidanak. \emph{This table is presented in its
    entirety in the electronic edition of the journal.  A portion is
    shown here for guidance regarding its form and content.}}
\label{phottable}
\end{deluxetable}

\begin{figure}[t!]
\plotone{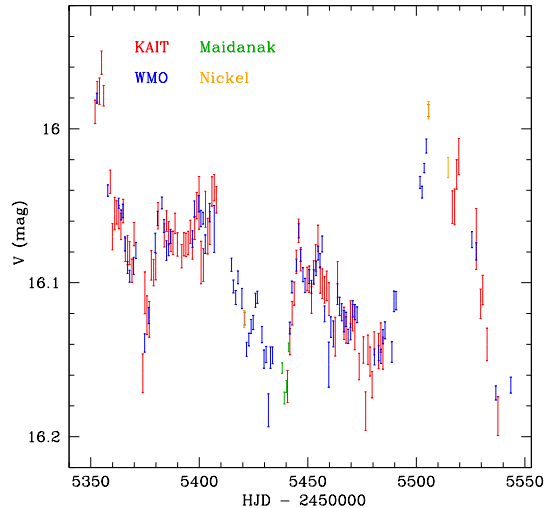}
\caption{$V$-band light curve of Zw 229-015.
\label{vlightcurve}}
\end{figure}

The final light-curve data for Zw 229-015 are listed in Table
\ref{phottable}.  Some observations taken in cloudy conditions had
very low signal-to-noise ratio (S/N), and we excluded any measurement
having a formal uncertainty greater than 0.03 mag.  We also discarded
measurements from a few images that suffered from very poor
flat-fielding or other problems.  The resulting light curve is
illustrated in Figure \ref{vlightcurve}.

The photometric uncertainties listed in Table \ref{phottable} are the
values returned by the photometry routine, which depend on
photon-counting statistics for the object and background as well as
on readout noise.  These are likely to be underestimates of the true
uncertainties, since additional errors can result from flat-fielding
irregularities, point-spread function variations, and other issues.
One way to estimate the actual photometric uncertainties is to examine
the magnitude fluctuations in the comparison-star light curves, which
are presumed to be intrinsically nonvariable.  Comparison stars 3, 4,
and 5 have $V$ magnitudes very similar to that of Zw 229-015.  For
each of these stars we examined the scatter of measurements about the
mean magnitude, and found standard deviations of 0.011--0.016 mag
(see Table \ref{compstarstable}), while the median value of the formal
photometric error on a single measurement for these stars is 0.006
mag.  This discrepancy shows that the photon-counting statistics are
usually not the dominant contributor to the error budget, except when
the images have very low S/N.  We address this issue by adding an
additional error term in quadrature to the photometric uncertainties
prior to carrying out the cross-correlation measurements, as described
in \S\ref{reverbsection}.

Our photometric aperture includes a substantial portion of the host
galaxy, including essentially the entire bulge and much of the bar and
disk.  We do not attempt to subtract off this host-galaxy starlight
contribution to the light-curve points, because our ground-based
imaging does not have sufficiently high resolution to decompose the
AGN point source and bulge into separate components.  Removing this
constant contribution of host-galaxy starlight would not affect the
measurement of the reverberation lag.

A condensed version of the $V$-band light curve was prepared for the
cross-correlation measurements by taking a weighted average of any
photometric data taken within 5 hours of each other.  This produces a
light curve with at most one data point per night.  The condensed
light curve was then converted from magnitudes to linear flux units.

\section{Spectroscopic Observations}

The spectroscopic monitoring program at the Lick 3~m Shane telescope
was carried out by several independent observing teams.  Unlike other
recent reverberation datasets that were obtained via long-duration,
dedicated campaigns of classically scheduled observing nights, we
chose to obtain synoptic monitoring of this single target by enlisting
the entire community of observers using the Kast spectrograph at Lick
during Summer and Fall 2010.  The Kast spectrograph is mounted at the
Cassegrain focus of the 3~m telescope for about half of each lunation,
and the spectrograph is used for a variety of programs including
supernova and transient follow-up observations, stellar spectroscopy,
and AGN surveys.  During each Kast observing night, the regularly
scheduled observers took a 1200~s exposure of Zw 229-015 to contribute
to this monitoring program.  This strategy enabled us to collect a
spectroscopic dataset with almost nightly sampling during each dark
run, except for weather-related losses.  The spectroscopic monitoring
began with one observation in May, and then continued with nightly
observations during each dark run (weather permitting) from June
through mid-December.  From June through September, nearly every night
was clear enough to observe, while October through December had more
substantial weather losses, as is typical for Mt.\ Hamilton (the site
of Lick Observatory).  Table \ref{spectable} gives a log of the Kast
observations for the 54 nights when useful data were obtained.  We
have omitted data from two nights when Zw 229-015 was observed through
very thick clouds, resulting in a S/N that was too low to be useful
for any measurements.

Each observing team used a different instrumental setup for the
spectrograph, designed for their primary science programs.  On the
blue side of the spectrograph, all observers used the 600-line grism,
which gives a dispersion close to 1.0~\AA\ pixel\per.  The spectral
range falling on the blue-side detector is controlled by adjusting the
CCD position via a manually movable stage, resulting in a fixed
wavelength coverage for the entire night's observations.  Most
observing teams set the CCD to a position that covered $\sim
3440$--5520~\AA, placing the \hbeta\ and [\ion{O}{3}]
$\lambda\lambda$4959, 5007 lines on the blue CCD.  However, on several
nights a bluer setup covering $\sim 3140$--5150~\AA\ or $\sim
3125$--5135~\AA\ was used, and with this setup the [\ion{O}{3}]
$\lambda$5007 line fell off the edge of the blue CCD, while the
$\lambda$4959 line was still present.

For the red side of the spectrograph, several gratings are available
and different gratings were installed on different nights.  Since we
could not ensure the same degree of consistency in the red-side
observations as was obtained on the blue side, the red-side
observations were a lower priority for this project.  On some nights
Zw 299-015 was not observed at all on the red side.  When it was
observed, gratings of 300, 600, and 1200 lines mm\per\ were used for
the red-side observations, with dispersions ranging from 1.2 to
4.6~\AA\ pixel\per.  In most cases, the observers used a d55 dichroic
to separate the blue and red beams of the spectrograph, while on a few
nights a mirror was used that directed all of the light to the blue
side.  When the 300 or 600 lines mm\per\ gratings were used, there was
usually a small region of wavelength overlap between the blue and red
spectra.  As a result of the heterogeneous nature of the red-side
data, this paper focuses primarily on the analysis of the blue-side
observations, which were all taken with the same grism and
homogeneously observed except for the variations in spectral range
described above.

With a single exception (the night of 2010 August 16), all
observations of Zw 229-015 were obtained with the spectrograph slit at
a position angle of PA = 56\fdg5.  This angle was chosen so as to
place a bright foreground star (marked with a square in Figure
\ref{wmoimage}) in the slit, in order to aid in centering the primary
target.  A 4\arcsec\ slit width was used for all exposures of Zw
229-015 and for the standard star BD+28\arcdeg4211.  A 1200~s
exposure of Zw 229-015 and a 60~s exposure of BD+28\arcdeg4211 were
taken on most nights.  On a few nights, the observers adjusted the
exposure time due to weather constraints or took more than one
exposure of the AGN.  The airmass was smaller than 1.2 for 72\% of the
spectroscopic observations of Zw 229-015, and smaller than 1.5 for all
but the final three exposures, when the object was observed at
airmasses between 1.5 and 2.0.  At the end of the observing season in
December, when the target was observed at relatively high airmass, the
fixed slit PA of 56\fdg5 was within 10--20\arcdeg\ of the optimal
parallactic angle \citep{filippenko1982} for the exposures, resulting
in minimal differential light losses.

Table \ref{spectable} also lists the ``image quality,'' measured as
the full width at half-maximum intensity (FWHM) of the spatial profile
of the Galactic foreground star in the spectroscopic exposure.  This
image quality is primarily a measure of atmospheric seeing, but on
some nights poor focus or occasional guiding errors added
significantly to the spatial profile width.  On clear nights, the
blue-side observations typically resulted in a S/N per pixel of $\sim
40$--45 in the extracted spectra at 4700~\AA.

Dome-flat exposures and observations of arc-line lamps were taken
during each afternoon.  Dome flats were usually observed through a
2\arcsec\ slit, since that slit width was used for the observers'
primary programs.  The difference in slit width between the AGN
exposures and dome-flat exposures did not have any significant impact
on the quality of the blue-side flattening, but on the red side it
left strong fringing residuals at wavelengths longer than $\sim
7000$~\AA.

\section{Spectroscopic Reductions and Measurements}

\begin{figure}
\plotone{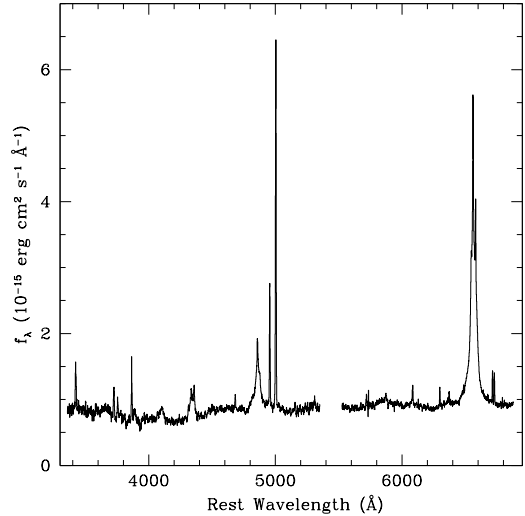}
\caption{Combined Kast blue-side and red-side spectrum of Zw 229-015, 
  from 2010-06-13.  On the red side, a 1200-lines mm\per\ grating was 
  used, giving a dispersion of 1.2~\AA\ pixel\per. 
\label{zw229spectrum}}
\end{figure}

Reduction of the Kast data followed standard procedures, including
overscan subtraction, flat-fielding, unweighted (nonoptimal)
extractions, and wavelength calibration using the line-lamp exposures.
An extraction width of 6\farcs5 was used.  Flux calibration and a
linear shift to the wavelength scale based on the wavelengths of
night-sky emission lines were then applied to the spectra.  The final
blue-side spectra were binned to a uniform linear scale of 1~\AA\
pixel\per.  Cosmic-ray residuals were removed from the reduced spectra
by interpolation over the affected pixels.  For nights when more than
one exposure was taken, the multiple exposures were combined into a
single reduced spectrum.  Error spectra were also extracted and
propagated through the full sequence of reductions.  Figure
\ref{zw229spectrum} illustrates the reduced spectrum from a single
observation.

We measured the centroid wavelength of the [\ion{O}{3}] $\lambda$5007
line in each spectrum where the line was available.  The median result
is $\lambda_\mathrm{obs} = 5143.37 \pm 1.17$~\AA, corresponding to an
emission-line redshift of $z = 0.0273 \pm 0.0002$, consistent with the
value from \citet{falco1999}.  The fairly large dispersion in the
[\ion{O}{3}] wavelength measurements is most likely a result of
variations in the centering of the AGN in the wide 4\arcsec\ slit in
each observation.

In order to intercalibrate the flux scales of the spectra and align
them to a consistent wavelength scale, we employed the scaling
procedure described by \citet{vgw1992}, which is based on the
assumption that the narrow [\ion{O}{3}] emission-line flux is
intrinsically constant over the timescale of the monitoring program.
This method applies small wavelength shifts and a multiplicative flux
scaling to each spectrum, as well as a convolution with a Gaussian
broadening kernel, and finds the values of the shift, kernel width,
and flux scaling factor that minimize the residual differences between
the individual spectrum and a reference spectrum over a small
wavelength region containing a narrow emission line and some
surrounding continuum.  According to \citet{vgw1992}, their scaling
method can yield accuracies of 5\% or better, but higher accuracy
($\sim 1$--2\%) can be obtained with the best-quality datasets
\citep[e.g.,][]{bentz2009,denney2010}.  The flux scale of the
reference spectrum was normalized by scaling it to match the mean
[\ion{O}{3}] $\lambda$5007 flux measured from eight nights that were
considered to be photometric by the observers.  For those eight
nights, we find a mean observed [\ion{O}{3}] $\lambda$5007 flux of
$3.96 \times 10^{-14}$ erg cm\persq\ s\per, with a standard deviation
of 4\%, as measured from the flux-calibrated spectra prior to applying
the scaling procedure.

When applying the \citet{vgw1992} procedure to the individual spectra,
we used the [\ion{O}{3}] $\lambda$4959 emission line as the flux
scaling reference line, instead of the $\lambda$5007 line.  The reason
for this modification was that 10 of the spectroscopic observations
used the blue-side CCD position that cut off the $\lambda$5007
emission line.  The wavelength range in common to all of the blue-side
exposures extended to 5130~\AA, including the $\lambda$4959 emission
line and about 20~\AA\ redward of the line.  By using the
$\lambda$4959 line as the flux reference, all of the spectra could be
treated identically in the flux scaling procedure.

To examine the accuracy of the spectral scaling, we measured the flux
in the $\lambda$4959 emission line from each spectrum after the
scaling was applied, along with its uncertainty from the propagated
error spectrum.  We then determined the normalized excess variance
$\sigma^2_\mathrm{x}$ of the $\lambda$4959 light curve, following
the definition used by \citet{nandra1997}:
\begin{equation}
\sigma^2_\mathrm{x} = \frac{1}{N\mu^2} \sum_{i=1}^N\left[(X_i -
\mu)^2 - \sigma_i^2\right], 
\end{equation}
where $N$ is the total number of observations, $\mu$ is the mean flux,
and $X_i$ and $\sigma_i$ are the individual flux values and their
uncertainties.  We find $\sigma_\mathrm{x} = 0.024$, indicating that
the residual errors in flux scaling contribute a spurious scatter to
the light curves at the $\sim 2$\% level.  These residuals are
primarily due to seeing variations (particularly from the worst-seeing
nights), miscentering of the target in the slit, and night-to-night
variations in the telescope and spectrograph focus.  Since Zw 229-015
exhibited dramatic, high-amplitude variability during the monitoring
period, this residual scatter only has a small effect on the
\hbeta\ light curve.

\begin{figure}
\plotone{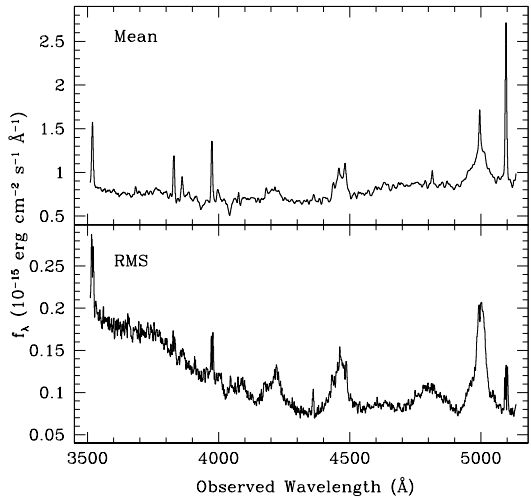}
\caption{Mean and RMS spectra.  The plot illustrates the wavelength
  range in common to all of the blue-side spectra over the course of
  the campaign.  Narrow emission-line residuals in the RMS spectrum
  become stronger toward the blue end due to errors in wavelength
  calibration and differences in spectral focus between nights, since
  the spectral scaling optimizes the alignment and resolution of the
  spectra based on the $\lambda$4959 emission line at the red end of
  the blue-side spectra.
\label{figmean}}
\end{figure}

Figure \ref{figmean} illustrates the mean and root-mean-square (RMS)
spectra based on the full dataset, constructed after applying the
spectral scaling procedure to the data. The RMS spectrum essentially
gives the standard deviation of flux values at a given wavelength, and
illustrates the relative degree of variability of different wavelength
regions of the spectrum.  The broad \hbeta, \hgamma, and \hdelta\
lines clearly stand out in the RMS spectrum, and the variable
continuum rises strongly toward the blue end of the observed
wavelength range.  In addition, a broad \ion{He}{2} $\lambda$4686 line
is evident in the RMS spectrum, although it is too weak to be visible
in the mean spectrum, where only the narrow component is clearly
detected.

The flux scaling of the red-side data is more difficult to establish
accurately.  On the red side, there are no strong and isolated narrow
emission lines, so the \citet{vgw1992} scaling routine cannot be
applied directly.  Additionally, on some nights the standard star was
not observed simultaneously on the blue and red sides, resulting in
flux scaling offsets when conditions were not photometric or the
seeing was poor.  In order to place the red-side data on a flux scale
that was approximately consistent with the blue-side data, we first
scaled each red spectrum by the scale factor used for the
corresponding blue spectrum.  This worked well for most nights, but
left obvious offsets between the blue-side and red-side flux scales
for a few nights.  For those nights, an additional scaling factor was
applied to the red-side data in order to bring the spectrum into
agreement with the corresponding blue-side spectrum over the region of
wavelength overlap or near-overlap.  (We note that in a dedicated
monitoring campaign with homogeneous observations, these flux scaling
problems can be mitigated by using a spectrograph setup with
substantial wavelength overlap between the red and blue sides, and by
observing standard stars simultaneously on both arms of the
spectrograph.)

\begin{figure*}
\begin{center}
\scalebox{0.6}{\includegraphics{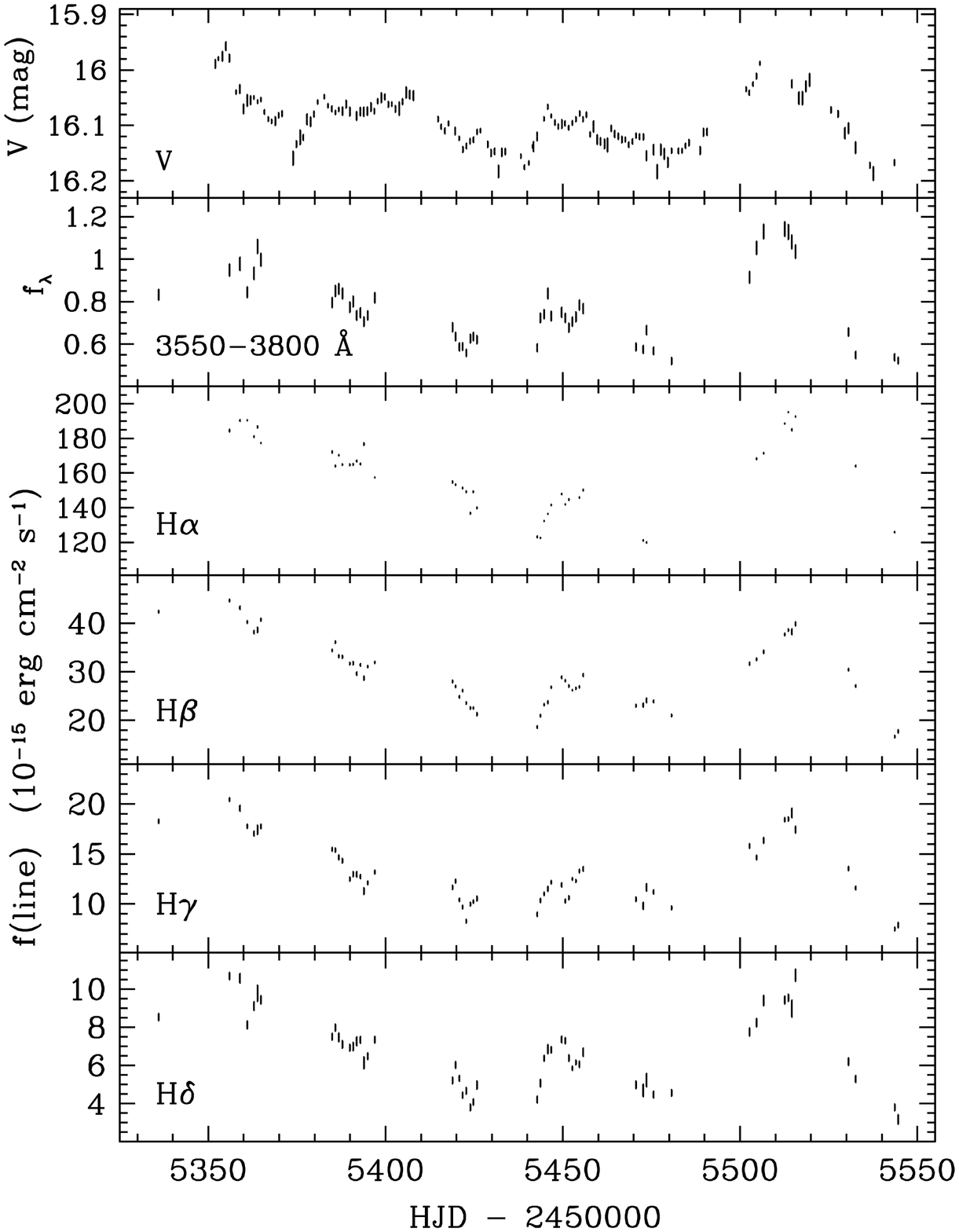}}
\caption{Light curves of the $V$-band continuum, the spectroscopic
  continuum over 3550--3800~\AA\ (in units of $10^{-15}$ erg cm\persq\
  s\per\ \AA\per), and the \hal, \hbeta, \hgamma, and \hdelta\
  emission lines.  The displayed $V$-band light curve is the
  ``condensed'' version in which measurements separated by less than 5~hr 
  have been averaged together giving a single data point per
  night.  The 3550--3800~\AA\ continuum light-curve points are plotted
  with error bars of $\pm 3\%$ as a rough approximation of the
  uncertainties from the spectral scaling procedure and slit losses,
  since these errors dominate over photon-counting uncertainties.
\label{alllightcurves}}
\end{center}
\end{figure*}

The \hbeta\ light curve was measured by first subtracting a local,
linear continuum underlying the line.  The continuum fitting windows
were set to 4860--4910~\AA\ and 5110--5130~\AA.  The \hbeta\ flux was
then measured by direct integration of the spectrum over
4930--5050~\AA.  We similarly measured the light curves of \hgamma\
and \hdelta.  The emission-line summation windows for \hgamma\ and
\hdelta\ were 4420--4500~\AA\ and 4175--4245~\AA, respectively, while
the continuum windows were 4300--4400~\AA\ and 4550--4600 \AA\ for
\hgamma, and 4100--4150~\AA\ and 4275--4325~\AA\ for \hdelta.  On the
red side, the emission-line window for \hal\ was 6600--6825~\AA, and
the continuum windows were 6500--6560~\AA\ and 6940--7020~\AA.  All of
these emission-line measurements include contributions from the narrow
components of the Balmer lines, which are presumed to be nonvariable
over the timescale of the monitoring project.  In the mean spectrum,
the narrow component of \hbeta\ contributes $\sim 10$\% of the total
\hbeta\ flux.  In addition, the \hgamma\ integration window includes
the [\ion{O}{3}] $\lambda$4363 line, and the \hal\ window includes
[\ion{N}{2}] $\lambda\lambda$6548, 6583.  We do not attempt to
decompose these emission blends into separate components.

\begin{deluxetable*}{llcccccc}
\tablecaption{Spectroscopic Observations and Measurements}
\tablehead{
  \colhead{UT Date} &
  \colhead{HJD} &
  \colhead{FWHM} &
  \colhead{S/N} & 
  \colhead{$f$(\hal)} & 
  \colhead{$f$(\hbeta)} & 
  \colhead{$f$(\hgamma)} & 
  \colhead{$f$(\hdelta)} \\
  \colhead{} &
  \colhead{$-2450000$} &
  \colhead{(arcsec)} &
  \colhead{} &
  \multicolumn{4}{c}{($10^{-15}$ erg cm\persq\ s\per)}
}
\startdata
2010-05-19 & 5335.995  & 2.2 & 43  &   \nodata         &    $ 42.39 \pm 0.24$  &    $ 18.27 \pm  0.20$ &   $ 8.54 \pm  0.18$  \\
2010-06-08 & 5355.978  & 2.1 & 45  &  $184.52 \pm0.62$ &    $ 44.69 \pm 0.25$  &    $ 20.44 \pm  0.20$ &   $10.69 \pm  0.19$  \\
2010-06-11 & 5358.954  & 2.9 & 37  &  $190.32 \pm0.30$ &    $ 43.19 \pm 0.32$  &    $ 19.56 \pm  0.26$ &   $10.56 \pm  0.25$  \\
2010-06-13 & 5360.985  & 2.5 & 40  &  $190.51 \pm0.23$ &    $ 40.25 \pm 0.27$  &    $ 17.76 \pm  0.22$ &   $ 8.12 \pm  0.21$  \\
2010-06-15 & 5362.928  & 2.7 & 40  &  $181.04 \pm0.33$ &    $ 38.19 \pm 0.29$  &    $ 17.04 \pm  0.24$ &   $ 9.11 \pm  0.22$  \\
\enddata
\tablecomments{The listed FWHM image quality is measured from
  the spatial profile of the bright star that fell on the spectrograph
  slit in each observation.  The listed S/N is the signal-to-noise ratio
  per pixel in the reduced blue-side spectrum at $\lambda = 4700$~\AA. 
  Measured fluxes include the contribution of blended
  narrow-line components. \emph{This table is presented in its
    entirety in the electronic edition of the journal.  A portion is
    shown here for guidance regarding its form and content.}}
\label{spectable}
\end{deluxetable*}

Figure \ref{alllightcurves} displays the Balmer-line light curves.
The \hbeta\ light curve is the cleanest of the four, because of the
proximity of the \hbeta\ line to the [\ion{O}{3}] line used as the
flux scaling reference.  The \hal\ light curve is substantially
noisier than that of \hbeta, despite the higher flux of \hal, due to
the issues in matching the flux scales of the red-side and blue-side
spectra.  Nevertheless, all of the lines clearly show the same pattern
of flux variations throughout the monitoring period.

The variability amplitude of the light curves can be quantified using
the statistics \rmax\ and \fvar\
\citep[e.g.,][]{kaspi2000,peterson2004,bentz2009}, where \rmax\ is the
ratio of maximum to minimum flux observed, and \fvar\ is calculated
according to
\begin{equation}
  \fvar = \frac{\sqrt{\sigma^2 - \langle\delta^2\rangle}}{\langle f
\rangle}, 
\end{equation} 
where $\sigma^2$ and $\langle \delta^2 \rangle$ are the variance and the
mean-square measurement uncertainty in the fluxes, and $\langle f
\rangle$ is the average flux in the light curve.  For the \hbeta\
light curve of Zw 229-015, we find $\rmax = 2.68 \pm 0.04$ and $\fvar =
0.23$.  By both of these measures, Zw 229-015 displayed strong
variability, exceeding the values of \rmax\ and \fvar\ observed for
the entire sample of 13 AGNs observed in the 2008 Lick AGN Monitoring
Project during a shorter 2-month campaign \citep{bentz2009}.

Figure \ref{alllightcurves} also displays the light curve of the
continuum region 3550--3800~\AA.  (We refer to this spectroscopic
$U$-band continuum measurement as the $U_s$ light curve.)  This is the
region most dominated by the AGN continuum in the blue-side spectra.
The spectroscopic continuum light curve is considerably noisier than
the $V$-band light curve, as expected due to the combination of
seeing-dependent slit losses, miscentering of the AGN in the slit, and
the nonparallactic slit orientation.  Nevertheless, the measurements
illustrate that the $U$-band continuum generally follows the same
pattern of variations as the $V$-band light curve, with a factor of
$\sim 2$ variability amplitude during the monitoring period.

\section{Reverberation Measurements}
\label{reverbsection}

\subsection{The \hbeta\ Emission Line}

\begin{figure}[b!]
\plotone{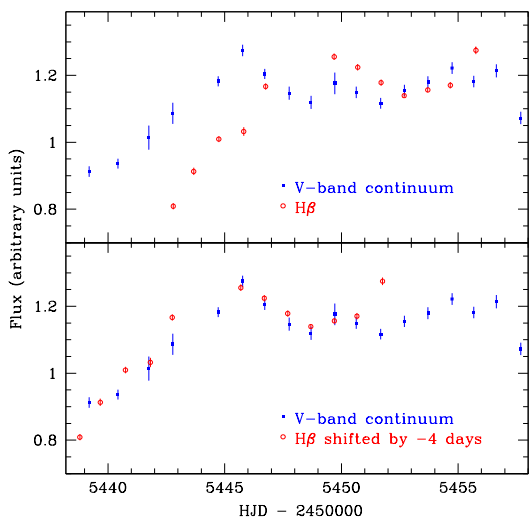}
\caption{Central portion of the $V$-band continuum and \hbeta\ light
  curves, illustrating the reverberation lag.  The light curves have
  been scaled to the same overall flux level, and a constant offset
  was first subtracted from the continuum so that its variability
  amplitude would match that of the \hbeta\ light curve.  In the lower
  panel, the \hbeta\ light curve is shifted by an offset of $-4$ days.
\label{lczoom}}
\end{figure}

Although it is difficult to see in Figure \ref{alllightcurves}, the
continuum and emission-line light curves do exhibit a small temporal
offset.  To demonstrate this more clearly, Figure \ref{lczoom}
illustrates the central portion of the $V$-band continuum light curve
(in flux units) and the \hbeta\ light curve, after rescaling to
approximately the same flux level.  Aligning the two light curves by
eye, it is evident that the \hbeta\ variations lag the continuum by
$\sim 4$ days.

To determine the reverberation lag quantitatively, we employed the
interpolation cross-correlation function (ICF) method
\citep{gaskellpeterson1987}, following the prescriptions described by
\citet{whitepeterson1994}, \citet{peterson2004}, and
\citet{bentz2009}.  The ICF method deals with unevenly sampled time
series by interpolating linearly between data points to obtain
temporally matched time series for the continuum and emission-line
light curves.  For a given trial value of the lag time $\tau$, the
continuum light curve is shifted forward in time by $\tau$ days, and
the cross-correlation is computed for the region of temporal overlap
between the two light curves.  The value of the cross-correlation
function (CCF) for a given lag is computed twice: first by
interpolating the shifted continuum light curve to the time steps of
the emission-line light curve, and then by interpolating the
emission-line light curve to the time steps of the shifted continuum
light curve.  The two interpolated versions of the CCF are then
averaged together to give the final CCF.  We computed the CCF for lag
values ranging from $-20$ to 30 days, in increments of 0.25 days.

Once the CCF is computed, the emission-line lag is determined in two
ways: the peak of the CCF defines \tpeak, while the centroid of the
CCF for points above 80\% of the peak value of the CCF gives \tcen.
The value of \tpeak\ and its uncertainty are quantized in units of the
sampling interval used for the cross-correlation, in this case 0.25
days.  Black hole masses from reverberation data are most often
determined using \tcen\ \citep{peterson2004}.

The final cross-correlation results and their uncertainties were
determined by following the Monte Carlo bootstrapping procedure
described by \citet{peterson1998}, \citet{welsh1999}, and
\citet{peterson2004}.  In this method, a large number of modified
realizations of the light curves are created and the cross-correlation
function is computed for each realization of a pair of continuum and
emission-line light curves.  Each simulated light curve is generated
by selecting $n$ points randomly from the original light curve, where
$n$ is the number of points in the original light curve and the random
selection allows a particular point to be chosen more than once.  If a
particular point is selected $m$ times, then its uncertainty is
reduced by a factor of $m^{1/2}$.  Then, the simulated light curve is
deviated by adding random Gaussian noise based on the uncertainty in
each data point.  The emission-line and continuum light curves are
each resampled $10^4$ times by this procedure, and for each pair of
simulated light curves the cross-correlation function is determined.
From these simulations, we build up distributions of values of \tcen\
and \tpeak.  The final adopted values of \tcen\ and \tpeak\ are
defined to be the median values from each distribution.  The $1\sigma$
uncertainty ranges on \tcen\ and \tpeak\ are given by the range of lag
values around the median that exclude the highest 15.87\% and lowest
15.87\% of lag values in the distribution.

When the continuum and emission-line light curves have very similar
observational cadences, the two interpolated versions of the CCF are
typically very similar in shape.  However, if the two light curves
have very different temporal sampling, then the two CCFs can differ
substantially.  As described by \citet{whitepeterson1994}, the two
interpolated versions of the CCF are usually averaged together to
obtain the final result, unless there is a strong reason to prefer one
interpolation over the other.  We tested whether this might affect the
lag measurements by measuring \tpeak\ and \tcen\ separately for the
two interpolated versions of the CCF, and comparing them with the
results obtained for the averaged CCF.  Although the two interpolated
CCFs differed somewhat in shape, the measured lags were very similar
in both cases, and the derived values of \tcen\ differed by only 0.18
days for the two separate CCFs.  Given this close agreement, we chose
to use the averaged CCF for our final results.

As discussed by \citet{welsh1999}, the reliability of
cross-correlation lag measurements can be improved by subtracting off
any long-term secular variations from the light curves that occur on
timescales much longer than the reverberation timescale.  In practice,
this is often accomplished by subtracting a simple linear fit from the
light curves.  We found that the measurements of \tpeak\ and
\tcen\ were virtually unchanged regardless of whether the light curves
were linearly detrended, and detrending with a quadratic fit (in an
attempt to remove the long-timescale curvature of the light curves)
only modified the measured lags by less than their $1\sigma$
uncertainties.  For our final measurements, we chose to use the
linearly detrended light curves.  We also tested to see whether the
cross-correlation measurements would be affected by removal of the
last few spectroscopic data points, during the period of poor weather
when the sampling was very sporadic, and found that the results for
\tpeak\ and \tcen\ again were consistent with our final adopted values
within the $1\sigma$ uncertainties.

As described above in \S\ref{photsection}, the photometric
errors listed in Table \ref{phottable} are likely to underestimate the
true uncertainties.  To account approximately for additional sources
of error, we added 0.01 mag in quadrature to all of the $V$-band
photometric errors prior to computing the cross-correlations.  We
found that this change had only a small effect on the
cross-correlation results: the derived value of \tcen\ was either
$3.85_{-0.90}^{+0.62}$ or $3.97_{-0.93}^{+0.71}$ days using the
original photometric errors or the expanded error bars, respectively.

\begin{figure}
\plotone{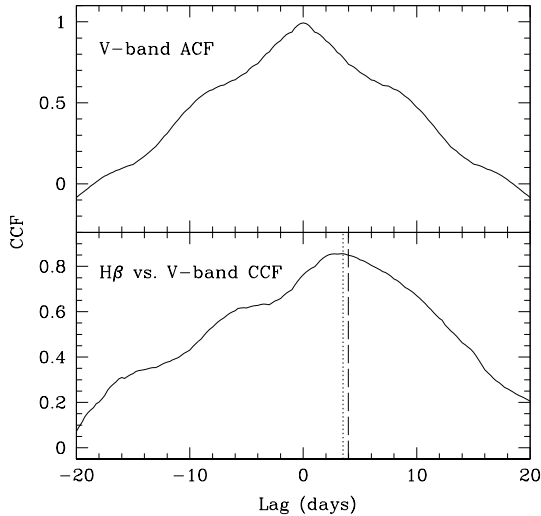}
\caption{\emph{Top:} Autocorrelation function of the $V$-band light
  curve.  \emph{Bottom:} Cross-correlation function of the \hbeta\
  and $V$-band light curves.  The dotted and dashed lines mark the
  values of \tpeak\ and \tcen, respectively.
\label{ccfs}}
\end{figure}

Our final measurement of the CCF gives observed-frame \hbeta\ lags of
$\tcen = 3.97_{-0.93}^{+0.71}$ days and $\tpeak =
3.50_{-0.75}^{+0.50}$ days, consistent with the general expectation of
a $\sim 4$ day lag from inspection of the light curves (as in Figure
\ref{lczoom}).  Converting to the rest frame of Zw 229-015, the lags
are $\tcen = 3.86_{-0.90}^{+0.69}$ and $\tpeak =3.41_{-0.73}^{+0.49}$
days.  All of the lag measurements are listed in Table \ref{lagtable}.
Figure \ref{ccfs} displays the autocorrelation function of the
$V$-band light curve and the \hbeta\ vs.\ $V$-band CCF.  The peak
value of the CCF is 0.86, indicating a fairly strong correlation
between the continuum and emission-line variations.

We also measured the \hbeta\ lag relative to the $U_s$-band light
curve and found observed-frame lags of $\tcen = 5.16_{-1.41}^{+1.27}$
and $\tpeak = 3.25_{-0.75}^{+2.75}$ days.  Within the substantial
uncertainties, these results are consistent with the lag measurements
carried out against the $V$ band.  Cross-correlating the $V$-band
light curve against the $U_s$ continuum, we find values of \tpeak\ and
\tcen\ that are consistent with zero within their $1\sigma$
uncertainties, so there is no significant detection of a lag between
continuum bands.  Higher-cadence observations would be needed in order
to search for any cross-band continuum lags
\citep[e.g.,][]{sergeev2005}.  In addition to the featureless
continuum of the AGN, the $U_s$ spectral region also includes emission
from the ``small blue bump,'' which contains both Balmer continuum and
\ion{Fe}{2} emission.  For NGC 5548, \citet{maoz1993} showed that the
small blue bump had a lag relative to the UV continuum that was
similar to the lag of Ly~$\alpha$, consistent with the expectation
that the small blue bump emission originates from the broad-line
region (BLR).  The nondetection of a lag between the $U_s$ and $V$
bands in Zw 229-015 suggests that either the small blue bump makes
only a small contribution to the $U_s$ band, or that the Balmer
continuum has an extremely short lag time relative to the primary
continuum.  The latter possibility would be consistent with the
decreasing lags measured for higher-order Balmer lines (see
\S\ref{balmerlags} below).

There is a small contribution to the $V$-band flux from the broad
\hbeta\ line, and also an extremely small contribution from \hal\ at
the red end of the Johnson $V$ filter passband.  This can potentially
bias the lag measurement, by adding a component to the continuum light
curve that has zero temporal shift with respect to the emission-line
light curve. Consequently, the lag measured from the cross-correlation
peak will tend to be lower than the true lag.  In order to assess the
magnitude of this effect, we carried out synthetic photometry on
Zw~229-015 using a combined blue-side and red-side spectrum
constructed from a night with good overlap between the blue and red
sides.  Using the IRAF SYNPHOT package, we calculated the $V$-band
magnitude of the spectrum twice: first on the original spectrum, and
then on a modified version of the spectrum in which the \hbeta\ and
\hal\ emission lines were completely removed by manually interpolating
over them.  The two synthetic $V$-band measurements differed by only
0.02 mag, indicating that the broad Balmer lines make only a very
small contribution ($\sim 2\%$) to the broad-band $V$ magnitude.

To obtain a quantitative estimate of the magnitude of this bias on the
lag measurement, we carried out a set of Monte Carlo simulations.  We
simulated a large number ($10^4$) of AGN light curves using the method
of \citet{timmerkoenig}, for a power-density spectrum of the form
$P(f) \propto f^{-2.5}$ normalized to a continuum RMS variability of
20\% over a 180-day period. A simple delta-function transfer function
with a lag of 4 days was used to model the emission-line response to
the continuum fluctuations.  The ``contaminated'' $V$-band light curve
was simulated by adding the emission-line light curve to the continuum
light curve, weighted so that 2\% of the total flux was from the
emission-line contribution.  With each set of simulated light curves
we measured the emission-line lag relative to the uncontaminated
continuum light curve, and relative to the continuum light curve with
the added emission-line contribution.

The simulations confirm that the bias in the lag measurements is very
small for Zw~229-015.  When the emission-line contribution to the
continuum light curve is at the observed 2\% level as in Zw 229-015,
the simulations show that the measurement of \tcen\ is biased on
average by only $-0.07$ days (with an RMS scatter of 0.05 days in the
bias for the full ensemble of simulations), relative to the
measurement obtained by cross-correlating against the pure continuum
light curve.  For Zw 229-015, the bias is much smaller than the
overall uncertainty in the lag measurement, but it can become
significant for higher levels of emission-line contamination.  We
carried out additional simulations, and found that if the
emission-line contribution to the broad-band light curve increases to
10\%, for example, the average bias in the \tcen\ measurement is
$-0.3$ days (for the same model parameters described above).  For
future reverberation campaigns that use broad-band photometry to
measure the continuum light curves, it will be important to carry out
simulations like these to assess the likely magnitude of this effect
on each individual object.  The size of the bias would depend on both
the emission-line equivalent width and on the target's redshift, since
higher redshifts would move the broad \hbeta\ line closer to the peak
transmission of the $V$ filter.  Alternatively, another approach would
be to subtract off the emission-line flux directly from the broad-band
light curve before measuring the cross-correlation lag.

\begin{deluxetable}{lcc}
\tablecaption{Lag Measurements}
\tablehead{
  \colhead{Measurement} &
  \colhead{\tcen\ (days)} &
  \colhead{\tpeak\ (days)}
}
\startdata
\hal\ vs.\ $V$    & $5.22_{-1.18}^{+0.83}$ & $4.75_{-0.50}^{+1.00}$ \\
\hbeta\ vs.\ $V$  & $3.97_{-0.93}^{+0.71}$ & $3.50_{-0.75}^{+0.50}$ \\
\hgamma\ vs.\ $V$ & $3.46_{-1.24}^{+0.80}$ & $2.50_{-0.75}^{+0.75}$  \\
\hdelta\ vs.\ $V$ & $1.20_{-1.69}^{+1.44}$ & $1.75_{-0.50}^{+0.75}$  \\
\hbeta\ vs. $U_s$ & $5.16_{-1.41}^{+1.27}$ & $3.25_{-0.75}^{+2.75}$  \\
$V$ vs. $U_s$     & $0.89_{-1.76}^{+1.44}$& $-0.25_{-0.50}^{+0.75}$ \\  
\enddata
\tablecomments{All lags are in the observed frame.  Rest-frame lags can
  be obtained by dividing by $1+z$, or 1.0275 in this case.  $U_s$ 
  denotes the $U$-band continuum measured from the blue-side spectra 
  over 3550--3800~\AA.
 }
\label{lagtable}
\end{deluxetable}

\subsection{Velocity-Resolved Measurements}

The behavior of the reverberation lag as a function of velocity across
broad emission lines is dependent on the kinematics of the BLR, and
recent ground-based programs have successfully obtained
velocity-resolved measurements of reverberation lags for \hbeta\ and
other lines in several AGNs \citep{kollatschny2003, bentz2009,
bentz2010a, denney2010}.  Our Zw 229-015 dataset is not ideally suited
for velocity-resolved reverberation measurements, because the \hbeta\
lag is fairly short and the monthly gaps in the spectroscopic light
curve limit the measurement accuracy for lags of individual velocity
segments of the \hbeta\ line.

\begin{figure}
\plotone{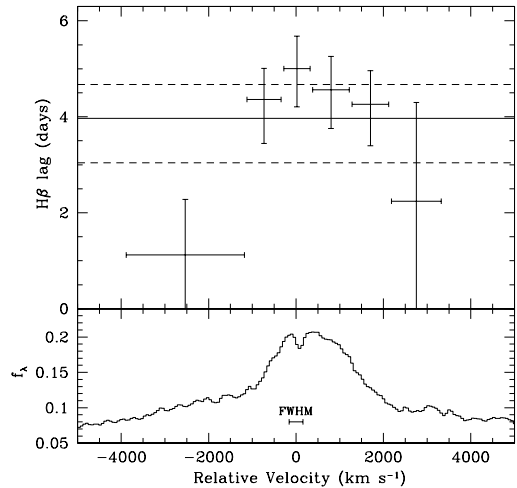}
\caption{Velocity-resolved reverberation results for \hbeta.  The
  lower panel shows the RMS line profile on a velocity scale, and the
  error bar illustrates the FWHM instrumental broadening of 313 \kms.
  The upper panel shows the value of \tcen\ measured in six velocity
  bins, and the horizontal error bars denote the size of each
  velocity bin.  The solid and dashed lines mark the value of \tcen\
  for the entire \hbeta\ line and its $1\sigma$ uncertainty range.
\label{vdelay}}
\end{figure}

Nevertheless, we are still able to obtain some rudimentary information
on the velocity dependence of the reverberation lag across the
\hbeta\ line profile.  Light curves were measured for several
independent velocity segments across the \hbeta\ line, and the
individual segment light curves were cross-correlated against the
$V$-band continuum.  The results are illustrated in Figure
\ref{vdelay}.  Six velocity segments were used for the final
measurements.  In the line wings, where the variability amplitude is
relatively low, the measurements have large uncertainties; the lags
measured for the blue-wing and red-wing segments are consistent with
zero within their $1\sigma$ uncertainties.  (We used a broader
velocity segment for the blue wing because subdividing the blue wing
into even smaller segments led to very poor cross-correlation results,
with some subsegments having negative lags.)

With these caveats in mind, the velocity-resolved behavior seen in
Figure \ref{vdelay} shows marginal evidence for a slightly longer lag
in the line core relative to the mean lag, while the lag in the line
wings is shorter.  A symmetric blue-red response in the velocity-delay
map, with shorter lags in the line wings, is qualitatively consistent
with Keplerian motion of the BLR clouds \citep[e.g.,][]{welsh1991},
but due to the large uncertainties in the velocity-resolved
measurements we refrain from drawing any specific conclusions about
the BLR kinematics.  The results do illustrate that Zw 229-015 is a
promising candidate for further velocity-resolved reverberation work.
Given the strong and rapid variability in this object, a high-cadence
monitoring campaign could obtain much better results, and a
reconstruction of the two-dimensional transfer function (i.e., the
distribution of lag time as a function of velocity) might be
obtainable \citep[e.g.,][]{bentz2010b}.

\subsection{Reverberation Lags of \hal, \hgamma, and \hdelta}
\label{balmerlags}

We measured cross-correlation lags for the other Balmer lines
following the same methods that were used for \hbeta, including the
linear detrending.  The Balmer-line reverberation lags follow the same
pattern found previously for other AGNs, where the \hal\ line has the
longest lag and higher-order Balmer lines have progressively shorter
lags \citep[e.g.,][]{kaspi2000, bentz2010a}.  The lag of \hdelta\ is
too small to be resolved significantly, and our measurement of \tcen\
indicates a $1\sigma$ upper limit of $\tcen<2.6$ days. As discussed
by \citet{bentz2010a}, this trend in Balmer-line lags is the result of
the differences in optical depths for the Balmer lines in the BLR, and
the same behavior is seen in BLR photoionization models
\citep{korista2004}.  For example, the \hal\ line has the highest
total optical depth through the BLR, and the observed \hal\ emission
originates primarily from larger radii in the BLR where the optical
depth is smaller.  Using the \tcen\ measurements, we find a ratio of
Balmer-line reverberation lags of
$\tau$(\hal):$\tau$(\hbeta):$\tau$(\hgamma) = 1.31:1.00:0.87, which is
within the range observed by \citet{bentz2010a} for AGNs having similar
emission-line lags of a few days.

\subsection{\hbeta\ Line Width}

Virial estimates of black hole masses based on reverberation-mapping
data employ the width of broad emission lines to give the typical
velocity for BLR gas.  Following procedures often employed for
reverberation-mapping data \citep[e.g.,][]{peterson2004}, we measured
the FWHM and the line dispersion (or second moment) of the \hbeta\
line from both the mean and the RMS spectra.  In both cases, a linear
continuum was fitted to regions on either side of the \hbeta\ line,
and then subtracted off.  The FWHM was measured directly from the
continuum-subtracted line profile, and the line dispersion \sigmaline\
(in velocity units) was calculated as
\begin{equation}
  \sigmaline^2 = \left(\frac{c}{\lambda_0}\right)^2 
 \left( \left[\sum \lambda_i^2 S_i\right] / \left[\sum
    S_i\right] - \lambda_0^2 \right),
\end{equation}
where $S_i$ is the flux density at wavelength bin $\lambda_i$, and
$\lambda_0$ is the flux-weighted centroid wavelength of the line
profile.  

While the narrow component of \hbeta\ essentially vanishes in the RMS
spectrum, it must be removed from the mean spectrum before the
broad-line width can be measured.  The narrow [\ion{O}{3}]
$\lambda$4959 emission line has a nearly Gaussian profile with FWHM =
6.6~\AA, and we used this as a model to subtract off the narrow
component of \hbeta.  To determine the final values and uncertainties
for the line widths, we followed the Monte Carlo method described by
\citet{peterson2004} and \citet{bentz2009}, creating randomly
resampled realizations of the mean and RMS spectra.  To construct each
realization, we randomly selected 54 spectra from our dataset, without
regard to whether an individual spectrum was already selected, and
then determined the mean and RMS spectra for that randomly sampled
set.  For a total of 1000 realizations, we measured FWHM and
\sigmaline\ for \hbeta, from both the mean and RMS spectra.  In each
realization, the borders of the continuum-fitting windows were shifted
randomly over a range of 10~\AA\ in order to account for the
line-width uncertainty resulting from the specific choice of continuum
regions.  The final value and uncertainty for each line-width
measurement (FWHM and \sigmaline) are taken to be the median value and
standard deviation determined from the full ensemble of simulations.
From the RMS spectrum, we measure FWHM = $2280 \pm 65$ \kms\ and
$\sigmaline = 1600 \pm 47$ \kms\ for \hbeta.  The mean spectrum gives
FWHM = $3360 \pm 72$ and $\sigmaline = 1645 \pm 16$ \kms\ for the
broad \hbeta\ line.  After removal of the narrow component from the
mean spectrum, the broad \hbeta\ line centroid in the observed frame
is $\lambda_0=4993.2$ \AA.  The redshift inferred from the broad
\hbeta\ line is therefore 0.0271, and there is no significant velocity
offset between the narrow [\ion{O}{3}] line and the broad \hbeta\
emission.

The observed line widths are affected by instrumental broadening in
the 4\arcsec\ wide slit, but only by a very small amount.  The
instrumental line width adds in quadrature to the observed line width
(as in $\sigma_\mathrm{observed}^2 = \sigma_\mathrm{intrinsic}^2 +
\sigma_\mathrm{instrumental}^2$, or similarly for the FWHM values).
We can obtain an upper limit to the instrumental broadening by
measuring the widths of arc-lamp lines observed through the
4\arcsec\ slit, since the arc lamps uniformly illuminate the entire
slit width while the AGN does not (at least under typical seeing
conditions).  From the \ion{Cd}{1} line at 5086~\AA, we find an upper
limit to the instrumental broadening of FWHM$_\mathrm{inst} <
7.7$~\AA, corresponding to an instrumental velocity width of
FWHM$_\mathrm{inst} < 454$ \kms\ or an instrumental dispersion of
$\sigmainst < 193$ \kms.  This instrumental broadening is small enough
that the maximum possible contribution of instrumental broadening to
the observed \sigmaline\ of the broad \hbeta\ line is only 0.9\% of
the observed line dispersion.

To estimate the likely value of the instrumental dispersion
\sigmainst, we followed the method described by \citet{bentz2009},
using an observation of [\ion{O}{3}] through a narrower slit to
determine the intrinsic width of the line.  The initial test spectrum
of Zw 229-015, taken on 2010 April 18, was observed through a
2\arcsec\ slit.  In that exposure, the [\ion{O}{3}] $\lambda$5007 line
has a raw FWHM of 5.44~\AA, and the 5086~\AA\ arc line has FWHM =
4.18~\AA.  Assuming that the AGN nearly fills the slit in the
2\arcsec\ observation (roughly the size of the seeing disk), this
implies an intrinsic line width of FWHM = 203 \kms\ for [\ion{O}{3}].
The instrumental broadening for the 4\arcsec\ slit can then be
estimated as the difference in quadrature between the observed width
of [\ion{O}{3}] in the 4\arcsec\ slit (FWHM = 373 km s\per) and the
intrinsic [\ion{O}{3}] width measured from the 2\arcsec\ observation.
This implies an instrumental contribution of FWHM$_\mathrm{inst} =
313$ \kms\ or $\sigmainst = 133$ \kms\ for the 4\arcsec\ slit,
consistent with the upper limits derived previously.

The corresponding corrections to the measured broad \hbeta\ line width
for this instrumental broadening are very small.  From the RMS
spectrum, the corrected \hbeta\ widths are FWHM = $2260 \pm 65$ \kms\
and $\sigmaline = 1590 \pm 47$ \kms, and in the mean spectrum the
corrected \hbeta\ widths are FWHM = $3350 \pm 72$ and $\sigmaline =
1640 \pm 16$ \kms.

\section{The Mass of the Black Hole}

Measurements of reverberation lag and broad-line width can be combined
to give an estimate of black hole mass, under the assumption that the
line width is primarily due to virial motion of broad-line gas
\citep[e.g.,][]{gaskell1988,kaspi2000,peterson2004}.  From the virial
theorem, we have
\begin{equation}
\mbh = f \frac{ (c \tau)(\Delta V)^2}{G},
\label{virialequation} 
\end{equation}
where $\tau$ is the reverberation lag and $(c \tau)$ gives the mean
radius of the BLR, $\Delta V$ is some measure of the broad-line width
in velocity units (typically either FWHM or the line dispersion
$\sigma$), and $f$ is a scaling factor which is dependent on the
geometry and kinematics of the BLR as well as our viewing angle.
Since these parameters of the BLR are unknown, it has become customary
to use a single value for $f$ that should ideally represent the mean
value of the virial normalization factor for an appropriate ensemble
of AGNs.  The adopted value of $f$ can be based either on some set of
assumptions about BLR kinematics \citep[e.g.,][]{kaspi2000}, or by
finding the normalization that brings reverberation-mapped AGNs into
best agreement with the local \msigma\ relation of quiescent galaxies
\citep{onken2004, woo2010}.

For consistency with the majority of recent reverberation work, we use
\sigmaline(\hbeta) measured from the RMS line profile as the measure
of $\Delta V$, and $c\tcen$(\hbeta) as the measure of BLR size; these
parameters have been shown to yield the most robust mass estimates
\citep{peterson2004}.  For the virial normalization factor, we use $f
= 5.25$, derived from the full available sample of
reverberation-mapped AGNs having measured stellar velocity dispersions
\citep{woo2010}.  Combining the rest-frame value $\tcen =
3.86^{+0.69}_{-0.90}$ with $\sigmaline = 1590 \pm 47$ \kms, we obtain
a ``virial product'' of $(c \tau)(\Delta V)^2/G= 1.91_{-0.46}^{+0.36}
\times 10^6$ \msun.  Applying the virial normalization factor $f=5.25$,
this gives an estimated black hole mass of $\mbh =
1.00_{-0.24}^{+0.19} \times 10^7$ \msun.  

We note that the quoted uncertainties on \mbh\ above only include the
propagated errors on the lag and line-width measurements, but not the
(unknown) systematic uncertainty on the applied value of $f$.  From
examination of the scatter of reverberation-mapped AGNs about the
best-fitting \msigma\ relation, the typical uncertainty in the value
of $f$ applied to any individual AGN is probably at the level of a
factor of $\sim 3$ \citep[e.g.,][]{onken2004, woo2010}, so the actual
error in the estimated value of \mbh\ is dominated by the uncertainty
in $f$ and not by the measurement errors in the \hbeta\ lag or width.
A long-term goal of reverberation mapping work is to use
velocity-resolved measurements to directly constrain the BLR geometry
and kinematics so that accurate black hole masses can be derived for
individual AGNs, but such analysis is beyond the scope of this work.

We can estimate the Eddington ratio \eddratio\ for Zw 229-015 using
the $V$-band continuum flux and a bolometric correction.  Without
high-resolution optical images or an accurate model of the stellar
population, we cannot directly determine the amount of starlight
contamination to the $V$-band magnitudes or spectroscopic continuum
fluxes, but a rough approximation will suffice to estimate \eddratio.
According to the extinction map of \citet{schlegel1998}, the Galactic
foreground extinction toward Zw 229-015 is $A_V = 0.24$ mag.  We make
the simplifying assumption that the extinction correction and
starlight contamination correction would roughly compensate for each
other, and take the rest-frame 5100~\AA\ flux density from the mean
spectrum as an estimate of the intrinsic (extinction-free) AGN flux
density.  We find $\lambda f_\lambda$(5100~\AA) $ = 4.2 \times
10^{-12}$ erg s\per\ cm\persq\ in the mean spectrum, corresponding to
$\lambda L_\lambda$(5100~\AA) $ = 7.1 \times 10^{42}$ erg s\per\ for a
luminosity distance of 119 Mpc (assuming WMAP-7 cosmological
parameters with $H_0=71$ km s\per\ Mpc\per; Larson \etal\ 2011).
Applying the same bolometric correction used by \citet{peterson2004}
of $\lbol = 9 \times \lambda L_\lambda$(5100~\AA), we obtain an
estimated $\lbol = 6.4 \times 10^{43}$ erg s\per, and $\eddratio
\approx 0.05$.  This value is within the typical range found for
nearby reverberation-mapped Seyferts \citep{peterson2004}.

The possible influence of radiation pressure on black hole masses
derived from reverberation mapping has been the subject of much recent
discussion \citep{marconi2008, netzer2009, marconi2009, netzer2010}.
The impact of radiation pressure on the motion of BLR clouds is likely
to be strongest for AGNs radiating at high luminosity, but Zw 229-015
has a rather modest Eddington ratio and any corrections for radiation
pressure should be relatively minor.  \citet{marconi2008} proposed a
modification to the virial equation, adding a term proportional to the
AGN continuum luminosity, as in
\begin{equation}
  \mbh = f \frac{ (c \tau)(\Delta V)^2}{G} + 
  g \frac{L_{5100}}{10^{44}~\mathrm{erg}~\mathrm{s^{-1}}},
\end{equation}
where $L_{5100} = \lambda L_\lambda$ at 5100~\AA, and where the
numerical value of $f$ differs from the value used to normalize the
``standard'' virial equation (Equation \ref{virialequation}).  Using
the sample of reverberation-mapped AGNs having measured stellar
velocity dispersions, they estimated best-fit values of $f = 3.1$ and
$\log g = 7.6$.  Applying this formalism to Zw 229-015, we obtain an
estimate of $\mbh \approx 8.8 \times 10^6$ \msun\ for Zw 229-015.
This value is lower than the pure virial estimate for \mbh, due to the
combination of the smaller $f$ value derived by \citet{marconi2008}
and the relatively low luminosity of Zw 229-015, but it differs from
the pure virial estimate by only 12\%, or less than $1\sigma$.  From
examination of the orbits of BLR clouds under the influence of
radiation pressure, \citet{netzer2010} proposed an alternative variant
of the virial equation as a new mass estimator, dependent on the AGN
continuum luminosity and broad-line FWHM.  Applying their mass
estimator (their Equation 18) and using FWHM(\hbeta) measured from the
mean spectrum, we obtain $\mbh \approx 1.25 \times 10^7$ \msun\ for Zw
229-015.  In either case, the corrections to \mbh\ for
radiation-pressure effects are not very large, and they are far
smaller than the overall uncertainty in the value of the virial
normalization factor $f$ for an individual AGN.

\section{Discussion and Conclusions}

This dataset was obtained using a somewhat unconventional strategy, in
which the spectroscopic observations were done on a nightly basis, but
only during the dark runs.  We find that this method can deliver
high-quality reverberation results, provided that a well-sampled
photometric light curve (not having large gaps) is also obtained.
Since there are many facilities that can deliver nightly,
queue-scheduled imaging observations without monthly gaps, this is now
a straightforward approach to measurement of long-term light curves
and reverberation lags for additional AGNs.  In terms of the
measurement accuracy for the \hbeta\ lag, our results for Zw 229-015
are similar in quality to what was obtained for several objects with
similar lag times of a few days during the 2008 Lick AGN Monitoring
Project \citep{bentz2009}, a program comprising 64 mostly consecutive
nights of spectroscopic observations along with nightly photometry.

Our results demonstrate that Zw 229-015 will be an excellent target
for a variety of future investigations.  Its strong and rapid
variability is noteworthy in comparison to many other objects having
similar black hole masses and luminosities \citep[e.g.,][]{bentz2009},
and it is very fortunate that such a nearby and highly variable AGN
happens to be in the \kepler\ field of view.  While the \kepler\ light
curve for Zw 229-015 is not yet available, we anticipate that it will
enable fundamentally new investigations into the properties of AGN
optical continuum variability.  There are only a few other
low-redshift Seyfert galaxies being monitored by \kepler, and although
their variability properties are currently unknown, it seems likely
that Zw 229-015 will be the most important object of this small
sample.  As such, it is crucial that Zw 229-015 continue to be
monitored by \kepler\ for the longest possible duration during the
mission.

Additional observations of Zw 229-015 can be used to measure the
galaxy's bulge properties, to place it on the correlations between
black hole mass and bulge luminosity or mass, and to add it to the
sample used to calibrate the relationship between BLR
radius and continuum luminosity following the methods described by
\citet{bentz2006}.  Similarly, a measurement of the bulge velocity
dispersion for Zw 229-015 can be used to add this galaxy to the
\msigma\ relation for reverberation-mapped AGNs \citep{nelson2004,
onken2004, woo2010}.  Our wide-slit observations are not well suited
to carrying out measurements of stellar kinematics, but stellar
absorption features including the \ion{Ca}{2} triplet absorption lines
are easily visible in the Lick spectra, so it should be
straightforward to measure the bulge velocity dispersion with suitable
new observations. Other observing programs for Zw 229-015 are
currently in progress, including near-infrared monitoring with
\emph{Spitzer} to measure the dust reverberation timescale.  We will
also revisit our \hbeta\ reverberation measurement when the \kepler\
light curve becomes available, since the sampling and precision of the
\kepler\ data may lead to an improved determination of the
emission-line lag.

To summarize our conclusions: we have measured a rest-frame
reverberation lag of $\tcen =3.86_{-0.90}^{+0.69}$ days for the
\hbeta\ line in Zw 229-015.  Combining this result with the broad
\hbeta\ line width, we obtain a virial estimate of $\sim 10^7$ \msun\
for the black hole in this galaxy, under the assumption of a virial
normalization factor of $f=5.25$.  The mass estimate does not change
significantly if we adopt the method of \citet{marconi2008} to obtain
a mass estimate corrected for the dynamical influence of radiation
pressure in the BLR.  The \kepler\ target Zw 229-015 is destined to be
one of the most important AGNs for investigations of the origin and
characteristics of optical continuum variability.  As recent
observations have begun to reveal connections between optical
continuum variability characteristics and black hole masses in AGNs,
this estimate of \mbh\ will be a valuable asset for future
interpretation of the \kepler\ light curve for Zw 229-015.

\acknowledgments

Research by A.J.B., M.L.N., and J.L.W. at UC Irvine has been supported
by NSF CAREER grant AST--0548198.  A.V.F.'s group is grateful for the
financial support of NSF grant AST--0908886, the TABASGO Foundation,
Gary and Cynthia Bengier, and the Richard and Rhoda Goldman Fund;
also, J.M.S. thanks Marc J. Staley for a Graduate Fellowship.  KAIT
and its ongoing operation were made possible by donations from Sun
Microsystems, Inc., the Hewlett-Packard Company, AutoScope
Corporation, Lick Observatory, the NSF, the University of California,
the Sylvia \& Jim Katzman Foundation, and the TABASGO Foundation.  The
West Mountain Observatory 0.9~m telescope has been supported by NSF
grant AST--0618209. M.D.J., C.D.L., E.G.H., and J.W.M. would like
thank the Department of Physics and Astronomy at Brigham Young
University for continued support of research efforts at the West
Mountain Observatory.  Tommaso Treu's research group at UCSB has been
supported by NSF CAREER grant NSF--0642621 and a Packard Fellowship.
Research by J.H.W.  has been supported by Basic Science Research
Program through the National Research Foundation of Korea funded by
the Ministry of Education, Science and Technology (2010-0021558).  We
thank Rick Edelson for discussions that motivated this project.  We
are grateful to the Lick Observatory staff for their assistance in
obtaining the spectroscopic observations, and to the staff of the
Maidanak Observatory for obtaining data for this project.  We also
thank Tabitha Buehler, Carl Melis, Dovi Poznanski, X. Prochaska, and
Ben Zuckerman for contributing observations.  This research has made
use of the NASA/IPAC Extragalactic Database (NED) which is operated by
the Jet Propulsion Laboratory, California Institute of Technology,
under contract with NASA.

 \end{document}